\documentclass[aps,twocolumn,groupedaddress,showpacs,prb,floatfix]{revtex4}
\usepackage{graphicx,rotating,subfigure,amsmath,amsfonts,amssymb,delarray}
\renewcommand{\vec}[1]{\boldsymbol #1}
\newcommand{\e}{\text{e}}

\def\l{\left}
\def\r{\right}
\def\12{\frac{1}{2}}

\begin{document}
\bibliographystyle{apsrev}


\title{Thermodynamics of multiferroic spin chains}


\author{J. Sirker}
\affiliation{Department of Physics and Research Center OPTIMAS, University of Kaiserslautern, D-67663 Kaiserslautern, Germany}
\affiliation{Max-Planck-Institute for Solid State Research, Heisenbergstr.~1,
  70569 Stuttgart, Germany}

\date{\today}

\begin{abstract}
  The minimal model to describe many spin chain materials with
  ferroelectric properties is the Heisenberg model with ferromagnetic
  nearest neighbor coupling $J_1$ and antiferromagnetic next-nearest
  neighbor coupling $J_2$.  Here we study the thermodynamics of this
  model using a density-matrix algorithm applied to transfer matrices.
  We find that the incommensurate spin-spin correlations - crucial for
  the ferroelectric properties and the analogue of the classical
  spiral pitch angle - depend not only on the ratio $J_2/|J_1|$ but
  also strongly on temperature. We study small easy-plane anisotropies
  which can stabilize a vector chiral order as well as the
  finite-temperature signatures of multipolar phases, stable at finite
  magnetic field. Furthermore, we fit the susceptibilities of LiCuVO$_4$,
  LiCu$_2$O$_2$, and Li$_2$ZrCuO$_4$. Contrary to the literature, we
  find that for LiCuVO$_4$ the best fit is obtained with $J_2\sim 90$
  K and $J_2/|J_1|\sim 0.5$ and show that these values are consistent
  with the observed spin incommensurability. Finally, we discuss our
  findings concerning the incommensurate spin-spin correlations and
  multipolar orders in relation to future experiments on these
  compounds.
\end{abstract}

\maketitle

\section{Introduction}
\label{Intro}
A number of spin-$1/2$ chain materials has recently been investigated
in much detail which show multiferroic behavior, i.e., an intricate
interplay of magnetic and electric
order.\cite{MasudaZheludev,ParkChoi,SekiYamasaki,DrechslerVolkova,EnderleMukherjee,BuettgenKrugvonNidda,SchrettleKrohns}
A microscopic model for the electric ordering based on a spin current
mechanism has been introduced in Ref.~\onlinecite{KatsuraNagaosa} and
seems to provide an understanding for most of the experimental
findings. Here the electric polarization $\vec{P}$ is related to
non-collinear spin-spin correlations, $\vec{P}\sim
\vec{e}_{ij}\times (\vec{S}_i\times\vec{S}_j)$, where $\vec{S}_i$ is a
spin-$1/2$ operator at site $i$ and $\vec{e}_{ij}$ a vector connecting
the chain sites $i$ and $j$. Later it has been shown based on a
Ginzburg-Landau theory \cite{Mostovoy} and symmetry arguments
\cite{KaplanMahanti} that this coupling between the polarization and
magnetization always exists independent of the crystal symmetry. To
obtain a non-collinear spin structure on a lattice without geometrical
frustration, a minimal model has to contain additional interactions
apart from the nearest neighbor Heisenberg interaction.  One
possibility are {\it frustrating} longer range interactions. For chain
materials consisting of edge sharing copper-oxygen plaquettes like
LiCuVO$_4$, LiCu$_2$O$_2$ or Li$_2$ZrCuVO$_4$ the next-nearest
neighbor interaction is particularly relevant leading to the minimal
model
\begin{equation}
\label{intro1}
H= \sum_r \l\{ J_1 (\vec{S}_r\vec{S}_{r+1})_{XXZ} +J_2(\vec{S}_r\vec{S}_{r+2})_{XXZ} -hS^z_r\r\} .
\end{equation}
Here $(\vec{S}_r\vec{S}_{r+i})_{XXZ} = S^x_rS^x_{r+i} + S^y_rS^y_{r+i}
+\Delta S^z_rS^z_{r+i}$ with $\Delta\leq 1$ being an exchange anisotropy.
For the edge sharing chains the nearest neighbor coupling is
ferromagnetic ($J_1<0$) while the next-nearest neighbor coupling is
antiferromagnetic ($J_2>0$). This is the case we want to study here.
The external magnetic field is denoted by $h$. In the classical
isotropic model ($\Delta=1$) without field the frustration leads to a
helical spin arrangement with a pitch angle $\phi =
\arccos(1/4\alpha)$ for $\alpha =|J_2/J_1|> 1/4$. As in the
classical model, the ground state of the quantum model (\ref{intro1})
is ferromagnetic for $\alpha < 1/4$. For $\alpha >1/4$ the ground
state is a singlet whose nature has not fully been clarified yet.\cite{WhiteAffleck96,ItoiQin2}
Based on a renormalization group treatment starting from two decoupled
Heisenberg chains ($J_1=0$), it has been predicted that any small
$J_1<0$ produces a finite but tiny excitation gap.\cite{ItoiQin2}
Numerically such a gap could not be resolved so far.

Due to the $SU(2)$ spin rotational symmetry only a quasi long-range
helical order (algebraically decaying) is possible in the isotropic
model (\ref{intro1}) without field. Whether the correlation functions
are indeed algebraically or instead exponentially decaying with a very
large correlation length as suggested in Ref.~\onlinecite{ItoiQin2} is
a question which we will not address here and which is not important
for the following discussions. One can in any case still define a
pitch angle $\phi$ by studying the spin-spin correlation functions
which, however, turns out to be substantially modified compared to the
classical case due to quantum fluctuations.\cite{BursillGehring} If
the $SU(2)$ symmetry is broken by applying either a magnetic field $h$
or by an anisotropy $\Delta<1$ then the {\it vector chirality}
\begin{equation}
\label{intro2}
\kappa^{(n)} = (\vec{S}_r\times\vec{S}_{r+n})^z = \frac{i}{2}(S^+_rS^-_{r+n} - S^-_rS^+_{r+n})
\end{equation}
can have a nonzero expectation value because this requires only the
breaking of the remaining $\mathbb{Z}_2$ symmetry. The vector
chirality as defined in (\ref{intro2}) is directly related to the spin
current $j_r = j_r^{(1)} + j_r^{(2)}$ which can be obtained from a
continuity equation $\partial_t S^z +\partial_r j =0$. The equation of
motion yields $\partial S^z_r = i[H,S^z_r]=-J_1(\kappa_r^{(1)}-
\kappa_{r-1}^{(1)})-J_2(\kappa_r^{(2)}- \kappa_{r-2}^{(2)})$ which
leads to the indentification
$j=J_1\kappa^{(1)}+2J_2\kappa^{(2)}$.\cite{HikiharaKecke}

It has been shown recently that a magnetic field can also stabilize
multipolar phases apart from a phase with chiral order, both in the
isotropic as well as in the anisotropic
case.\cite{HeidrichMeisnerMcCulloch,SudanLuescher,HikiharaKecke,FurukawaSato}
Such multipolar phases are characterized by short-range transverse
spin correlations $\langle S^+_0S^-_r\rangle$ while correlations
functions $\langle\underbrace{S^+_0S^+_1\cdots}_{\tiny \mbox{n times}}
\underbrace{S^-_rS^-_{r+1}\cdots}_{\tiny \mbox{n times}}\rangle$ are
algebraically decaying in an $n$-polar phase.

In this article we want to investigate how the spin-spin correlations
of the $J_1$-$J_2$ model (\ref{intro1}) are affected not only by
quantum but also by thermal fluctuations. At the end we want to
relate our numerical results with various experiments on edge sharing
copper-oxygen chains. A very powerful numerical method to study the
thermodynamics of one-dimensional quantum chains directly in the {\it
  thermodynamic limit} is the density-matrix renormalization group
applied to transfer matrices (TMRG). Here the Hilbert space is
truncated to a small fixed number of states while the temperature is
lowered successively. In the calculations presented here we will
typically retain $N=60-200$ states both in the system and the environment
block. We will present data for temperatures where the algorithm seems
to be converged which is judged by comparing results obtained for
different numbers of states. For a detailed description of the TMRG
algorithm the reader is referred to
Refs.~\onlinecite{Peschel,SirkerKluemperPRB,SirkerKluemperEPL,GlockeKluemperSirker_Rev}.
In a previous TMRG study the isotropic $J_1$-$J_2$ model was
investigated, however, correlation functions were not
calculated.\cite{LuWang}

The manuscript is organized as follows: In Sec.~\ref{J1J2} we study
the isotropic model without magnetic field paying special attention to
the evolution of the incommensurabilities with temperature. In
Sec.~\ref{aniso} we consider the experimentally relevant case of a
small easy-plane anisotropy. In Sec.~\ref{finiteField} we investigate
signatures of multipolar phases, which are stable for finite magnetic
field, at finite temperatures. Finally, in Sec.~\ref{Materials}, we
relate our numerical results to data for three experimentally well
studied compounds (LiCuVO$_4$, LiCu$_2$O$_2$, Li$_2$ZrCuVO$_4$). In
particular, we fit the susceptibilities using model (\ref{intro1})
which allows us to extract the parameters $J_1$ and $J_2$. We also
discuss neutron scattering experiments which give access to the pitch
angle and suggest future experiments regarding the realization and
observation of multipolar orders. The last section is devoted to a
brief summary and some conclusions.

\section{Thermodynamics of the $J_1$-$J_2$ chain}
\label{J1J2}
\begin{figure}
\includegraphics*[width=0.99\columnwidth]{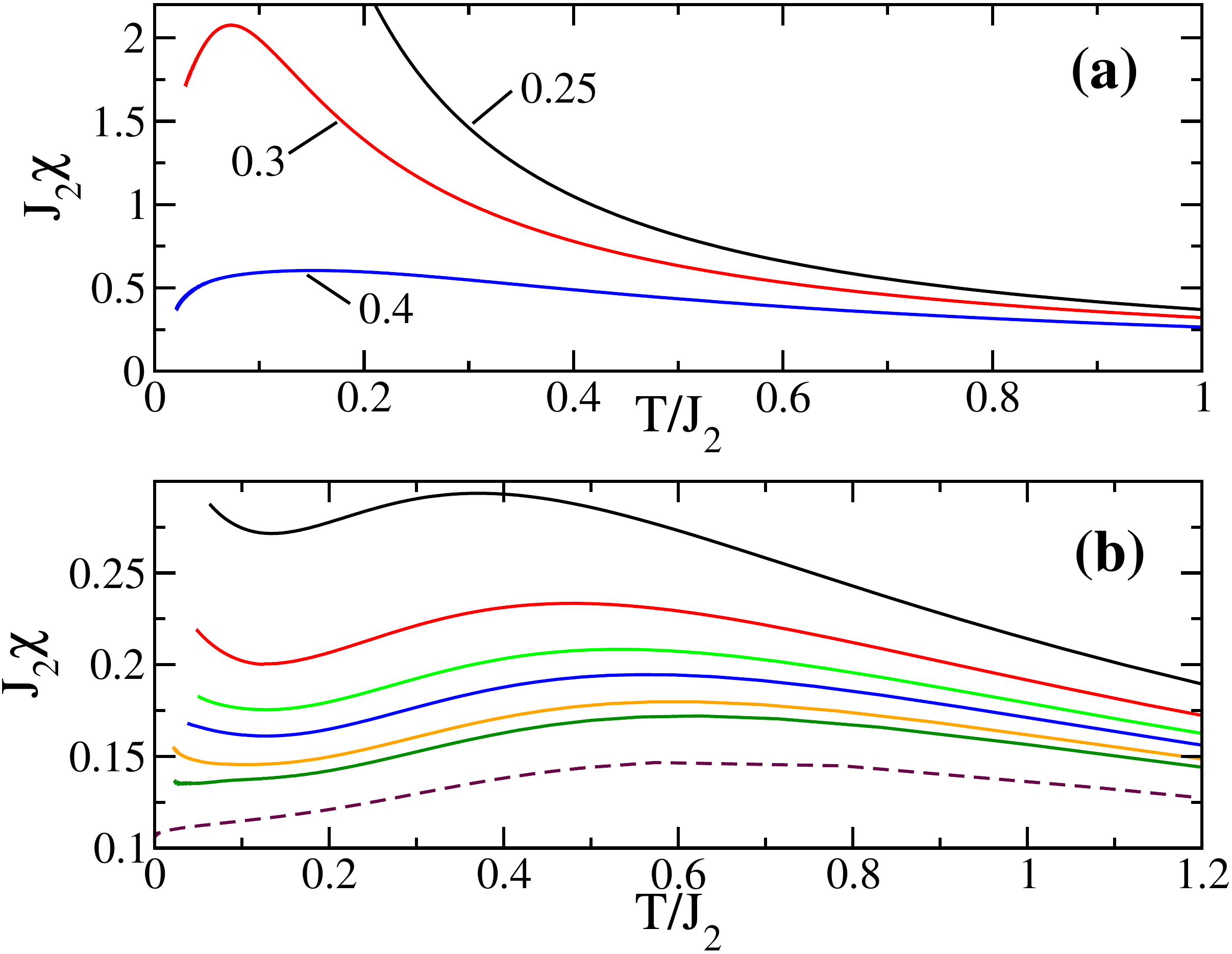}
\caption{(color online) (a) Susceptibilities near the quantum critical
  point $\alpha =1/4$. (b) Susceptibilities for $\alpha=2.0,\, 1.6,\,
  1.2,\, 1.0,\, 0.8,\, 0.6$ (solid lines from bottom to top). For
  comparison, the result for the nearest-neighbor Heisenberg chain
  obtained by Bethe ansatz is shown (dashed line).}
\label{J1J2.fig1}
\end{figure}

For the isotropic case, $\Delta=1$, without magnetic field it is known
that there is a quantum critical point at $\alpha = 1/4$ separating a
ferromagnetic phase for $\alpha < 1/4$ from a phase with a singlet
ground state. A tiny gap for $\alpha >1/4$ has been
predicted,\cite{ItoiQin2} however, even if this gap really exists it
will only show up at extremely low temperatures. In the temperature
range accessible numerically, the susceptibility $\chi$ for
$\alpha\to\infty$ smoothly approaches the result for the
nearest-neighbor Heisenberg chain known from Bethe ansatz
\cite{Kluemper_HB} as is shown in Fig.~\ref{J1J2.fig1}.
At the critical point, $\chi(T)$ will show a power-law
divergence.\cite{HaertelRichter} A thorough investigation of the
critical properties using the TMRG algorithm will be presented
elsewhere.\cite{RichterDrechslerSirker}

\begin{figure}
\includegraphics*[width=0.99\columnwidth]{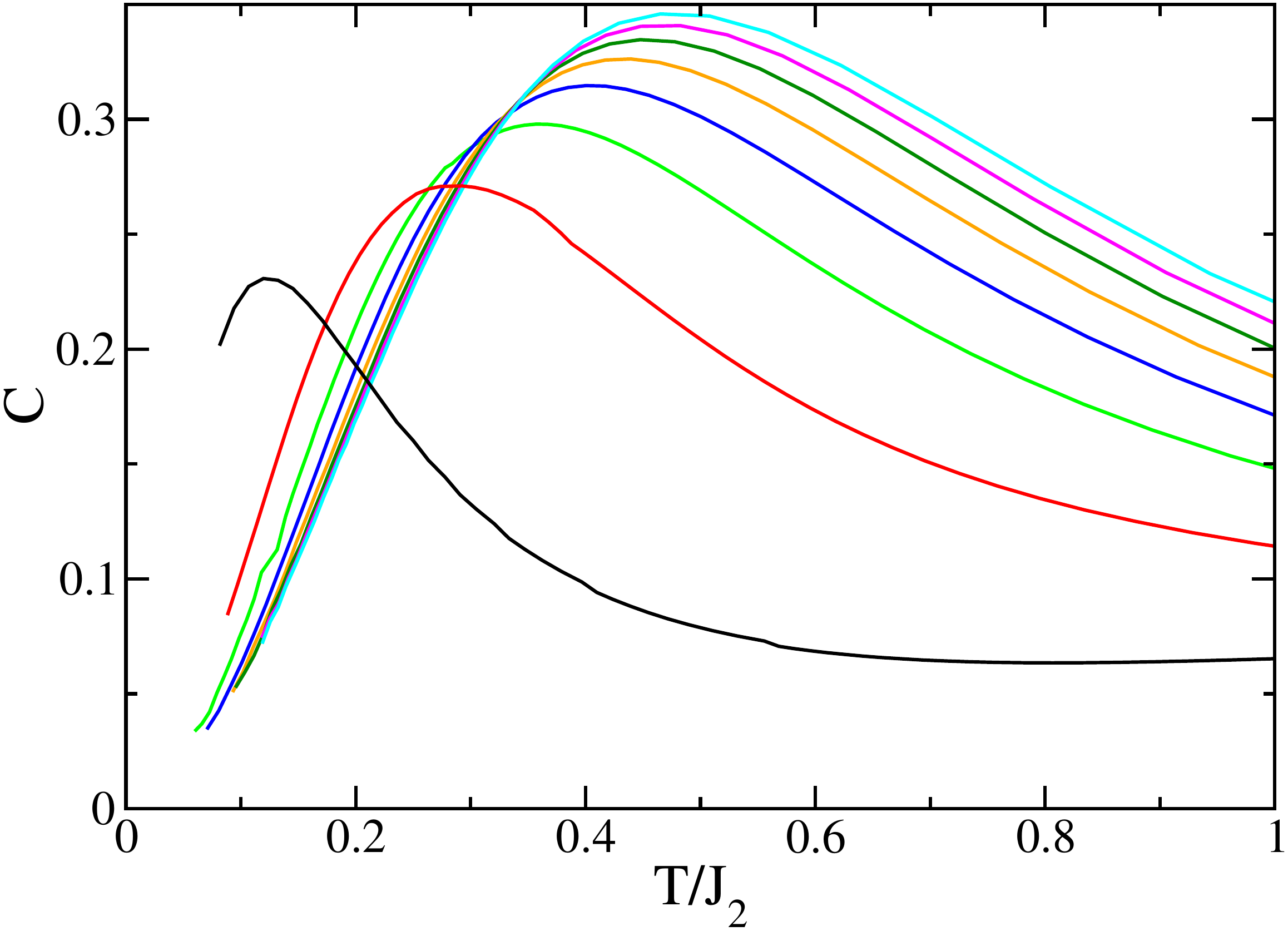}
\caption{(color online) Specific heat for $\alpha=0.4,0.6,\cdots,2.0$ (from
  bottom to top).}
\label{J1J2.fig2}
\end{figure}

Numerically, we can also obtain the inner energy $e$ and the free
energy $f$ (see Ref.~\onlinecite{GlockeKluemperSirker_Rev}). This
gives us the entropy $s=(e-f)/T$ and the specific heat $C=T\partial
s/\partial T$ using a numerical derivative.  The results for various
$\alpha$ are presented in Fig.~\ref{J1J2.fig2}.
For all frustration parameters we find a broad maximum which shifts
monotonically to larger temperatures $T/J_2$ with increasing $\alpha$.

Next, we want to study various correlation functions. At finite temperatures
all correlation functions will be exponentially decaying and we can expand any
two-point correlation function of an operator $O_r$ as 
\begin{equation}
\label{corr}
\langle O_0 O_r\rangle -\langle O_0\rangle\langle O_r\rangle = \sum_{j=0}^\infty  M_j \e^{-r/\xi_j}\cos(k_j r) \; .
\end{equation}
Here $M_j$ is a matrix element, $\xi_j$ a correlation length, and
$k_j$ the corresponding wave vector. Within the TMRG algorithm both
$\xi_j$ and $k_j$ are determined by the ratio of the leading to
next-leading eigenvalues of the transfer matrix.
\cite{Peschel,GlockeKluemperSirker_Rev} For large distances $r\gg 1$
the correlation function is dominated by the largest correlation
length and the corresponding wave vector which we will denote as
$\xi\equiv \xi_0$ and $k\equiv k_0$ in the following.

\begin{figure}
\includegraphics*[width=0.99\columnwidth]{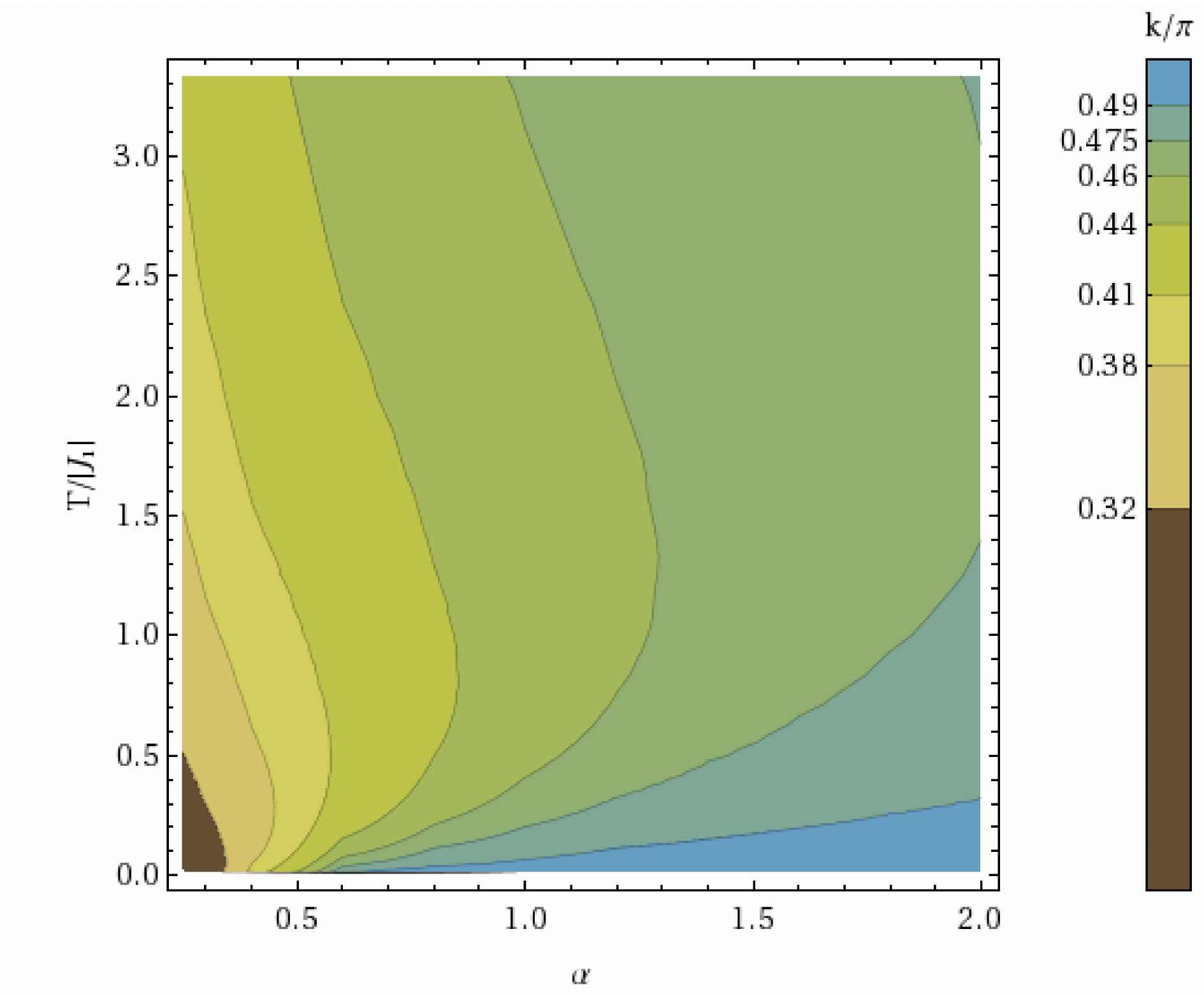}
\caption{(color online) The wave vector $k$ of the spin-spin
  correlation function depends not only on $\alpha$ but also on
  temperature. With increasing temperature, $k$ first increases for
  $\alpha$ close to $\alpha_c$ while it decreases for larger
  $\alpha$.}
\label{J1J2.fig3}
\end{figure}

\begin{figure}
\includegraphics*[width=0.99\columnwidth]{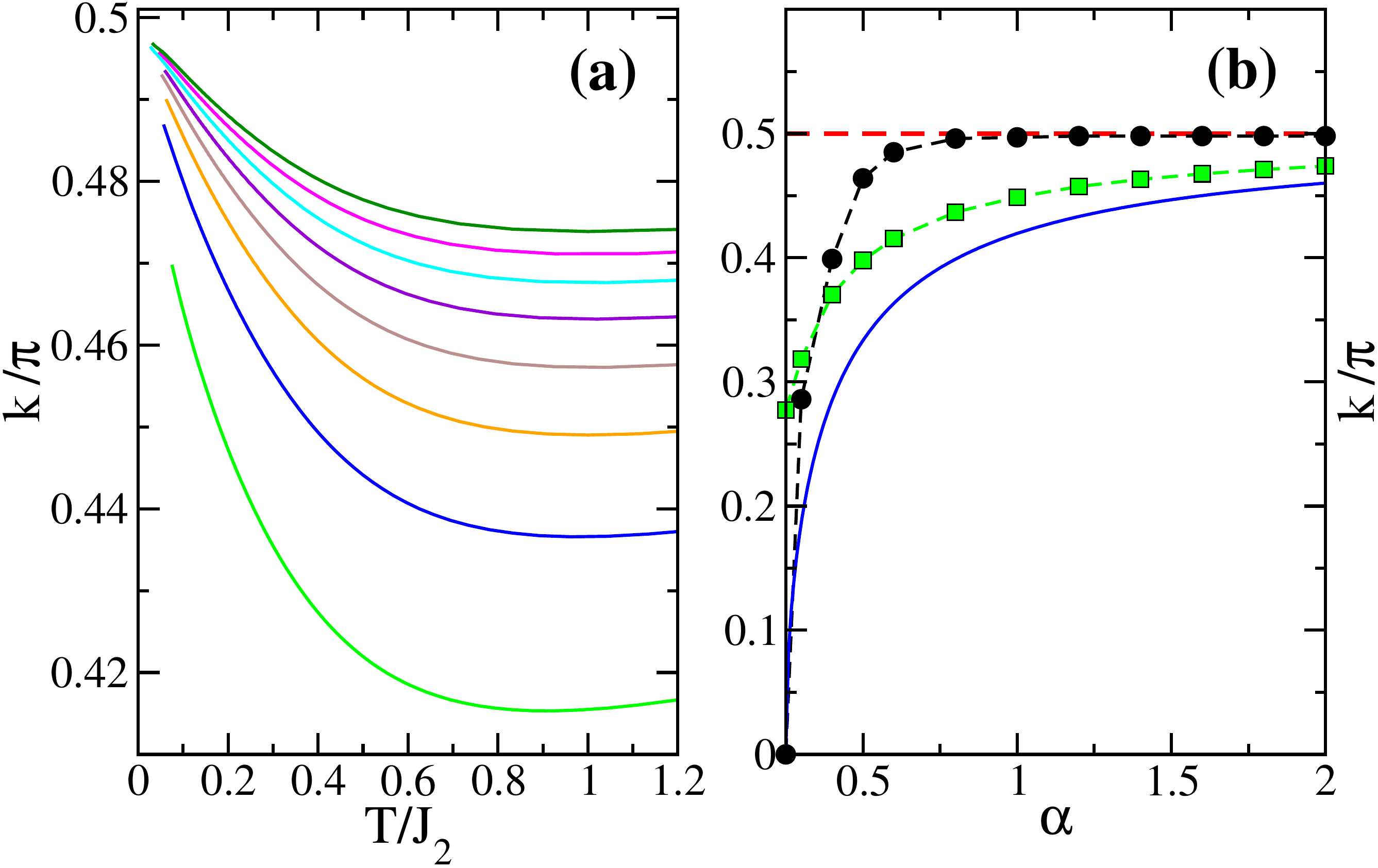}
\caption{(color online) (a) $k$ as a function of temperature for
  $\alpha=0.6,0.8,\cdots,2.0$ (from bottom to top). (b) Extrapolated
  $k$ (pitch angle) for zero temperature (circles), and $k$ at
  $T/J_2=1.0$ (squares). The dashed lines are a guide to the eye. For
  comparison, the classical pitch angle $\phi=\arccos(1/4\alpha)$
  (solid line) is shown.}
\label{J1J2.fig4}
\end{figure}

Of particular interest is the evolution of the wave vector $k$ for the
spin-spin correlation $\langle\vec{S}_0\vec{S}_r\rangle$. This can be
seen as the quantum mechanical analogue of the pitch angle of the
classical spiral. In Fig.~\ref{J1J2.fig3} it is shown that $k$ does
not only depend on the frustration $\alpha$ but also shows a strong
dependence on temperature, in particular, for values of $\alpha$ close
to the critical point $\alpha_c=1/4$.
Contrary to the classical case, the pitch angle at low temperatures is very
close to $\pi/2$ for $\alpha\gtrsim 0.6$ as is shown in more detail in
Fig.~\ref{J1J2.fig4}(a). 
This means that for these frustration values the spin structure
develops an almost perfect (short-range) four-site periodicity at low
temperatures.  In Fig.~\ref{J1J2.fig4}(b) the extrapolated $k$ value
(pitch angle) for zero temperature is shown as a function of $\alpha$.
Our result is in good agreement with an earlier DMRG
calculation.\cite{BursillGehring} We also find that with increasing
temperature the pitch angle becomes again more classical for a
frustration $\alpha\gtrsim 0.5$ (see squares in
Fig.~\ref{J1J2.fig4}(b)).

\begin{figure}
\includegraphics*[width=0.99\columnwidth]{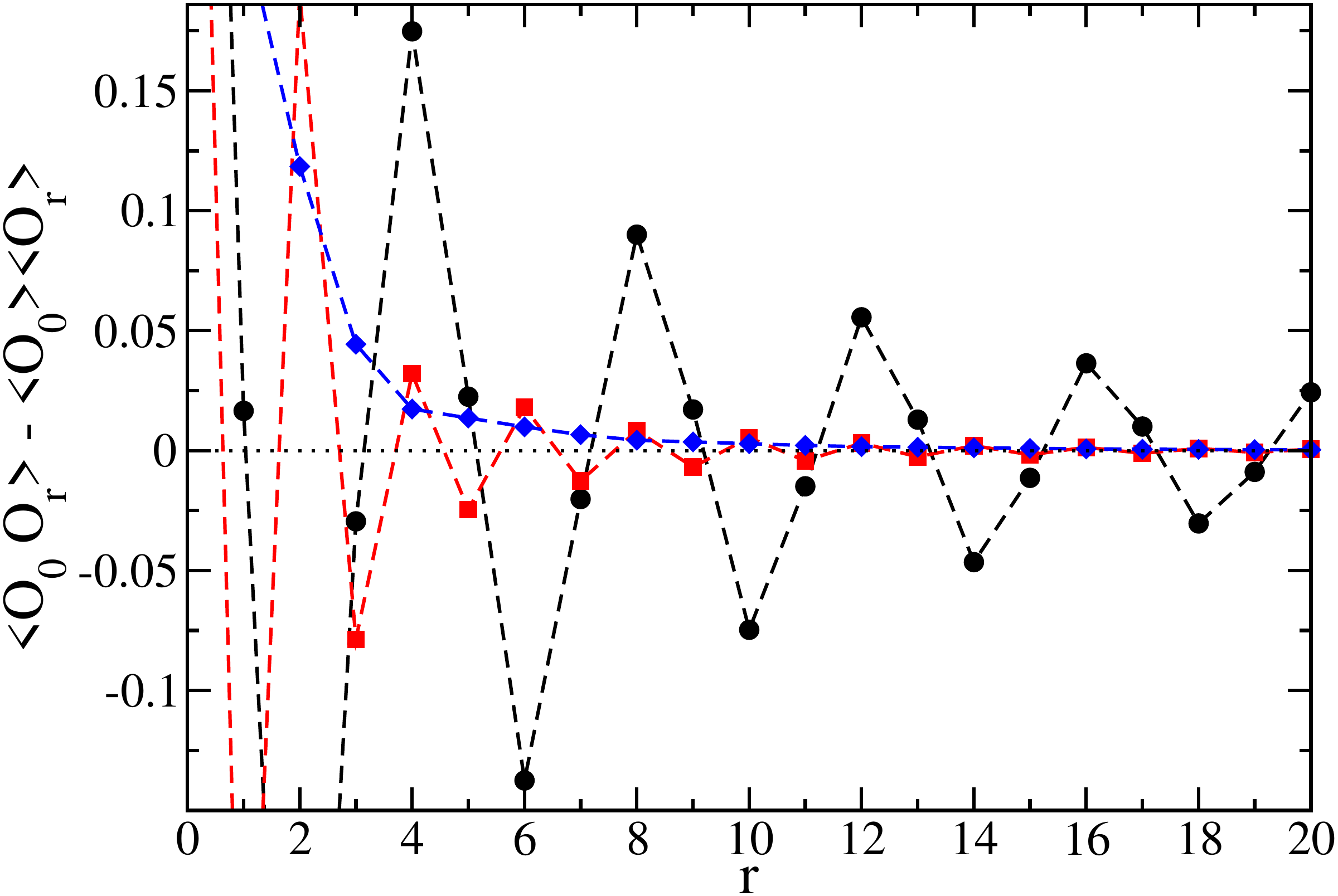}
\caption{(color online) Various correlation functions for $\alpha=2$
  and $T/J_2=0.05$. The circles denote the spin-spin
  ($O_r=\vec{S}_r$), the squares the dimer
  ($O_r=\vec{S}_r\vec{S}_{r+1}$), and the diamonds the chiral
  correlation function ($O_r=\vec{S}_r\times\vec{S}_{r+1}$). The lines
  are a guide to the eye.  }
\label{J1J2.fig5}
\end{figure}

One might think that the strong trend towards a formation of a
four-site periodic structure with increasing $\alpha$ could be related
to a concomitant dimerization. However, as is shown exemplarily in
Fig.~\ref{J1J2.fig5}, this is not the case.
While substantial dimer and chiral correlations with comparable
correlation lengths do exist, both correlation lengths are about a
factor $2$ smaller than the spin-spin correlation length.

\begin{figure}
\includegraphics*[width=0.99\columnwidth]{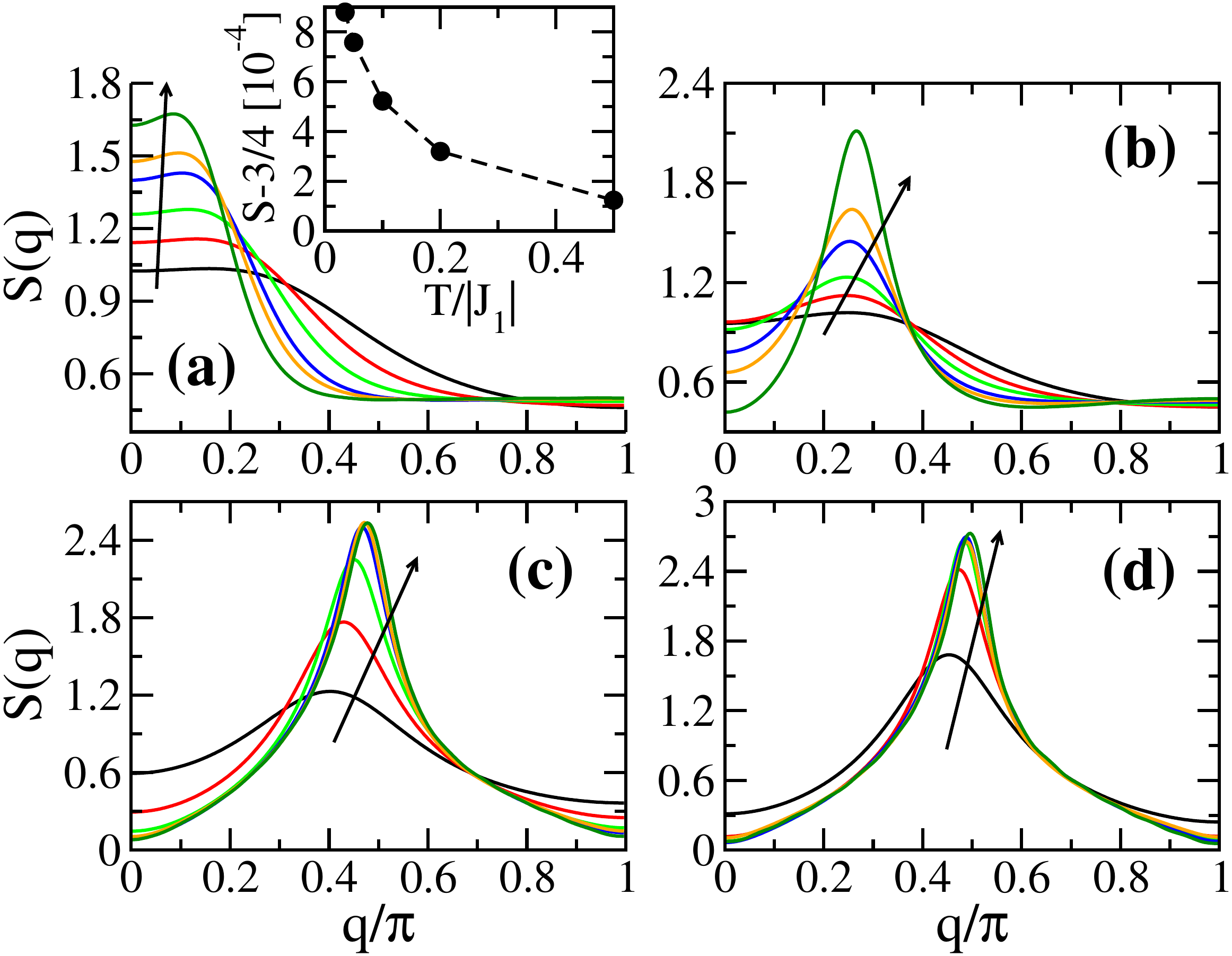}
\caption{(color online) The static spin structure factor $S(q)$ for
  temperatures $T/|J_1| = 0.5, 0.2, 0.1, 0.05, 0.035, 0.02$ (along the arrow
  direction) with (a) $\alpha = 1/4$ (quantum critical point), (b) $\alpha = 0.3$, (c)
  $\alpha = 0.6$, and (d) $\alpha = 1.0$. The inset of (a) shows the
  deviation from the sum rule $S\equiv \sum_q S(q) = 3/4$. 
}
\label{J1J2.fig6}
\end{figure}

Another way of defining a quantum analogue of the classical pitch angle is the
possibility to study at which wave vector $q$ the static spin structure factor
\begin{equation}
\label{S_of_q}
S(q)= \frac{3}{4} + 2\sum_{r=1}^\infty \cos(q\, r)\langle \vec{S}_0 \vec{S}_r\rangle
\end{equation}
is peaked. This definition of the pitch angle is equivalent to the
definition in (\ref{corr}) if the correlation function is dominated by
the leading correlation length, i.e., $\xi_0 \gg \xi_j$ for $j\geq 1$.
This is the case for all frustration parameters shown in
Fig.~\ref{J1J2.fig6} except for the critical value $\alpha_c=1/4$
(Fig.~\ref{J1J2.fig6}(a)).
Here the structure factor is very flat near $q\sim 0$ meaning that in
the expansion (\ref{corr}) comparable correlation lengths exist with a
wave vector $k=0$ and with wave vectors which have small
incommensurate values. The accuracy of the numerical data can be
checked by calculating the sum rule. For all $\alpha$ and all
temperatures the sum rule is fulfilled with deviations of the order
$10^{-3}$ to $10^{-4}$ as is exemplarily shown in the inset of
Fig.~\ref{J1J2.fig6}(a).

\section{The anisotropic case}
\label{aniso}
In the edge-sharing spin chain compounds substantial exchange
anisotropies exist. For LiCuVO$_4$, for example, the $g$ tensor has
been measured by ESR and an exchange anisotropy of the order of $6\%$
percent has been estimated.\cite{KrugvonNiddaSvistov} This anisotropy
is expected to pin the spins to the $ab$ plane and this is indeed
observed in experiment.\cite{EnderleMukherjee,SchrettleKrohns}
Theoretically, one expects that an $XXZ$-type anisotropy
($\Delta\lesssim 1$) enhances chiral correlations. In this case even
long range chiral order ($\langle \kappa^{(n)}\rangle\neq 0$) is
possible at zero temperature in the purely one-dimensional model
because only the remaining $\mathbb{Z}_2$ symmetry has to be broken.
In previous numerical studies the anisotropic model (\ref{intro1}) has
already been investigated, the results, however, have been
contradictory.\cite{SommaAligia,FurukawaSato} In
Ref.~\onlinecite{SommaAligia} a large region of the phase diagram has
been found to be occupied by a dimer phase whereas spin liquid phases
occur for very small and large values of $\alpha$. In
Ref.~\onlinecite{FurukawaSato} a dimer phase and spin liquid phases
have again been found but in addition also a chiral phase for $\alpha
\gg 1$. Both studies were based on exact diagonalization. In a very
recent density-matrix renormalization group study the anisotropic
model at finite magnetization was
investigated.\cite{HeidrichMeisnerMcCulloch} A dimer phase was not
found to be stable, instead it was shown that for most parameters a
chiral phase or a spin liquid phase (SDW$_2$ phase) occur.

\begin{figure}
\includegraphics*[width=0.99\columnwidth]{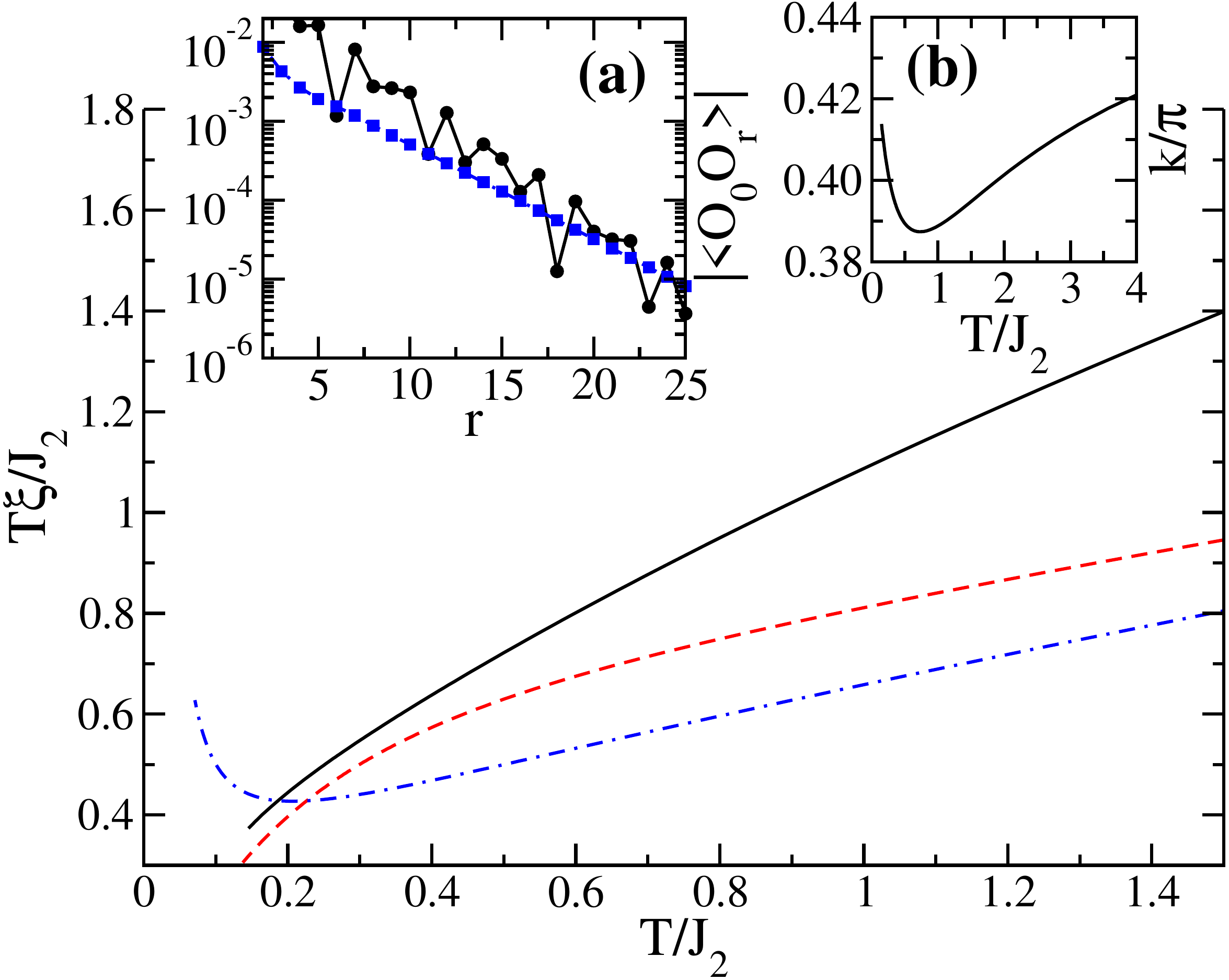}
\caption{(color online) The longitudinal spin (solid line), dimer
  (dashed line), and chiral correlation length (dot-dashed line) for
  $\alpha=0.4$ and $\Delta=0.8$. At low temperatures the chiral
  correlation length diverges stronger than $1/T$ indicating
  long-range order at zero temperature. (a) The longitudinal spin-spin
  and the chiral correlation function at $T/J_2=0.125$ showing that
  chiral correlations dominate. (b) The wave vector $k$ of the
  longitudinal spin-spin correlation function.}
\label{J1J2.fig7}
\end{figure}

We will first concentrate on one value of anisotropy, $\Delta=0.8$.
This is larger than what is expected for the edge-sharing cuprate
chains. However, this way the effects of the anisotropy are more
obvious and the results should qualitatively be similar to those for
realistic values for these materials. We will study the case of zero
field to see whether a dimer or a chiral phase occur.  In
Fig.~\ref{J1J2.fig7} results for various correlation lengths at
$\alpha=0.4$ are shown.
For correlation functions which decay algebraically at zero
temperature we expect that the correlation length diverges as $\xi\sim
1/T$. Studying $\xi T$ as in Fig.~\ref{J1J2.fig7} thus separates long
and short ranged correlations. In this case we see that the chiral
correlation length diverges stronger than $1/T$ indicating that for
these parameters we do have long range chiral order at zero
temperature. While the dimer is indeed larger than the chiral
correlation length over a wide temperature range, the situation is
reversed at low temperatures. Hence a dimer phase can only possibly
occur if also interchain couplings are present which might stabilize
such an order at intermediate temperatures. That the chiral
correlations indeed dominate at low temperatures is shown exemplarily
for $T/J_2=0.125$ in Fig.~\ref{J1J2.fig7}(a) where the chiral and
longitudinal spin-spin correlation functions are compared. Here the
wave vector $k$ of the longitudinal spin-spin correlation function is
incommensurate and again shows a strong dependence on temperature
(Fig.~\ref{J1J2.fig7}(b)).

\begin{figure}
\includegraphics*[width=0.99\columnwidth]{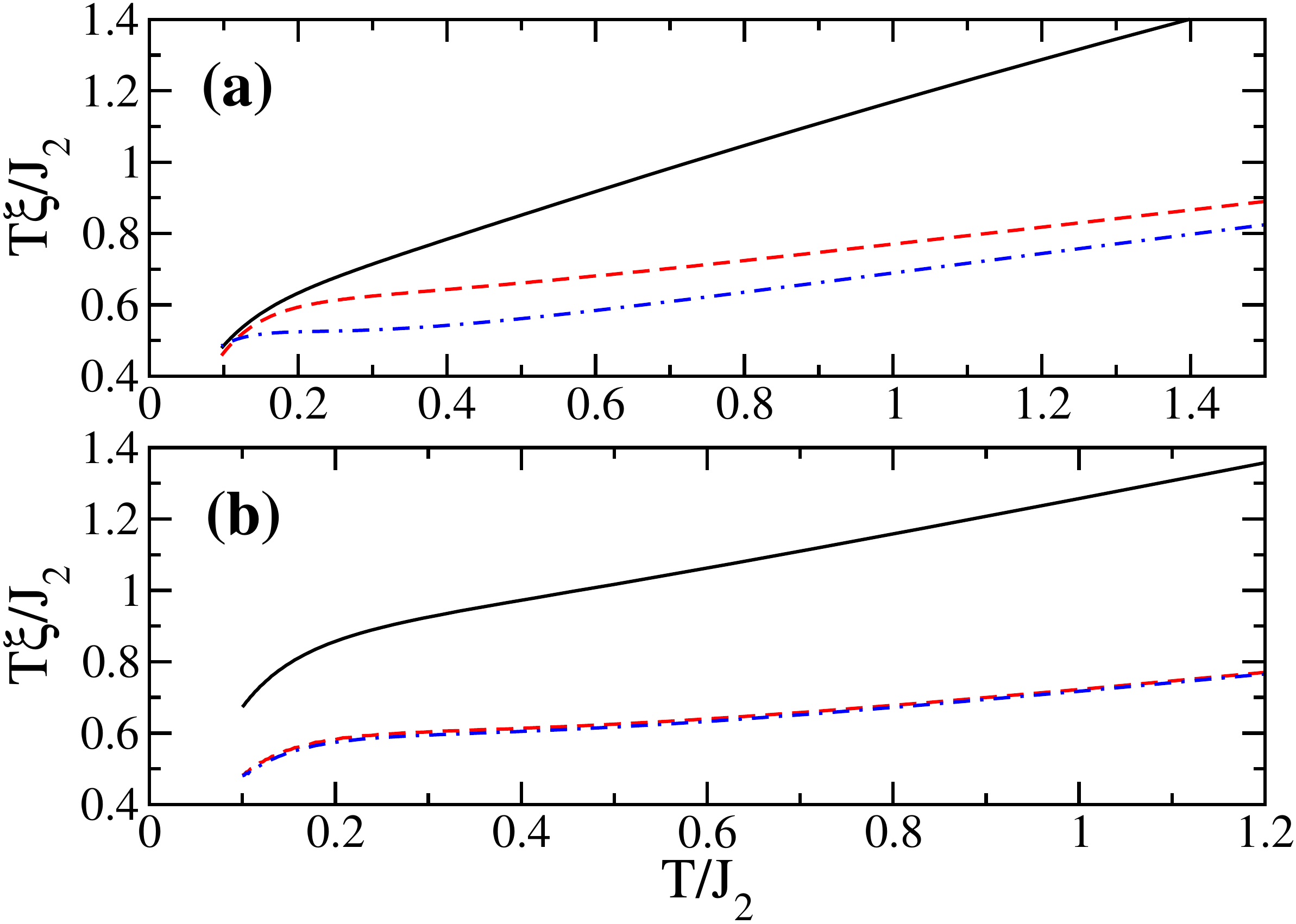}
\caption{(color online) The longitudinal spin (solid line),
  dimer (dashed line) and chiral (dot-dashed line) correlation
  lengths for $\Delta=0.8$ with (a) $\alpha=0.6$ and (b)
  $\alpha=2.0$.}
\label{J1J2.fig8}
\end{figure}

For $\alpha=0.6$, shown in Fig.~\ref{J1J2.fig8}(a), the chiral
correlation length is again dominant at the lowest temperatures which
are accessible numerically, however, there is no indication for long
range chiral order at zero temperature. Remarkably, all three
correlation lengths are of very similar magnitude at low temperatures.
This suggest that we are very close or at the phase transition from a
phase with long range chiral order at smaller values of $\alpha$ to a
phase which most likely has spin liquid character with algebraically
decaying correlation functions. 
Indeed, at $\alpha=2.0$ shown in Fig.~\ref{J1J2.fig8}(b) the spin-spin
correlation length is clearly dominant with oscillations $k\approx
\pi/2$ at low temperatures pointing to an SDW$_2$ phase. We conclude
that if a dimer phase exist at all for $\Delta =0.8$ it has to be
confined to a very narrow range of frustration parameters
$0.4<\alpha<0.6$. In Ref.~\onlinecite{SommaAligia} an unidentified
phase was found for $0.3\lesssim\alpha \lesssim 0.45$ with symmetries
as expected for the chiral phase. For $\alpha \gtrsim 0.7$, on the
other hand, a spin liquid phase was predicted with a dimer phase in
between these two phases. In our TMRG calculations we never find
dominant dimer correlations at low temperatures and our results
suggest that a dimer phase might not be stable at all but instead a
direct phase transition from the chiral to a SDW$_2$ phase at
$\alpha\approx 0.6$ occurs.

\section{Finite magnetic field: Multipolar phases}
\label{finiteField}
In the multiferroic spin chain compounds a magnetic field can be used
to switch the electric polarization. A sufficiently strong field
induces a flop of the spins from the easy-plane spiral to a spiral
perpendicular to the applied magnetic field. According to the spin
current mechanism this also leads to a rotation of the electric
polarization.  Such a switching of the ferroelectric polarization by
an applied magnetic field has been observed in LiCuVO$_4$
\cite{SchrettleKrohns} as well as in LiCu$_2$O$_2$.\cite{ParkChoi}
Numerical studies of the $J_1$-$J_2$ model have shown that small
magnetic fields stabilize the chiral order while multipolar phases
become stable at intermediate field strengths before the system
ultimately becomes fully polarized for large
fields.\cite{HikiharaKecke,SudanLuescher,FurukawaSato}

As already eluded to in the introduction, an algebraic decay of the
$\langle\underbrace{S^+_0S^+_1\cdots}_{\tiny \mbox{n times}}
\underbrace{S^-_rS^-_{r+1}\cdots}_{\tiny \mbox{n times}}\rangle$
correlation function is expected in an $n$-polar ($n\geq 2$) phase
while the transverse spin correlation function is gapped. For the
oscillations of the longitudinal spin correlation function we expect
in general
\begin{equation}
\label{multipolar}
k_{T\to 0} = \frac{\pi}{n}(1-2m)
\end{equation}
where $m$ is the magnetization and is directly related to the shift of
the Fermi points. For zero field, $n=2$ would correspond to the
$4$-site periodic spin structure discussed previously.

\begin{figure}
\includegraphics*[width=0.99\columnwidth]{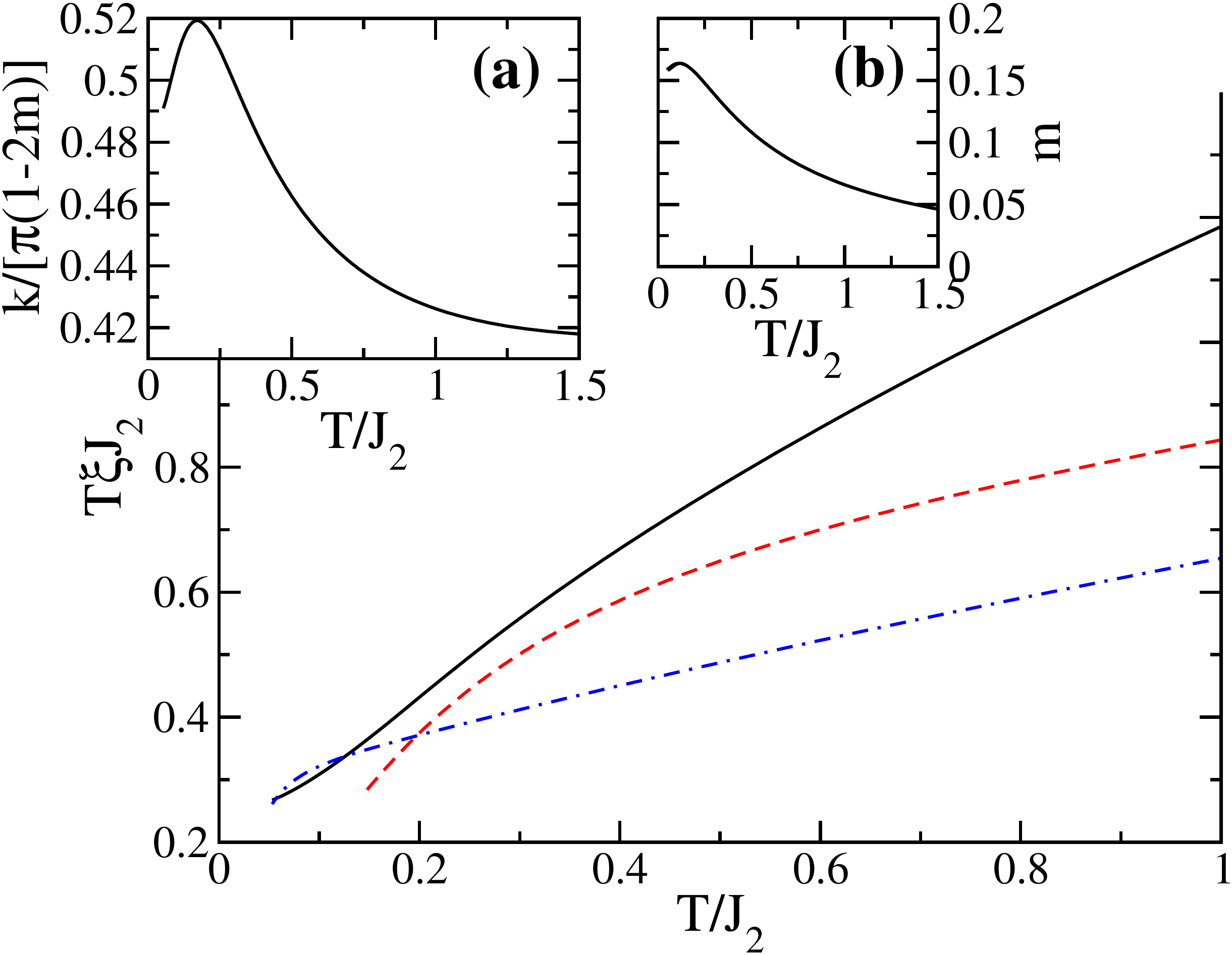}
\caption{(color online) The longitudinal spin (solid line),
  dimer (dashed line) and chiral (dot-dashed line) correlation
  lengths for $\alpha=0.4$ with $\Delta=1$ and $h=0.25\, J_2$. (a) The
  wave vector $k$ of the spin-spin correlation. (b) The magnetization
  $m$ as a function of temperature.}
\label{J1J2.finite_field.fig1}
\end{figure}

An particular interesting case is $\alpha=0.4$ with $h=0.25\, J_2$
shown in Fig.~\ref{J1J2.finite_field.fig1}. Here we find that the
longitudinal spin correlation length is largest at high
temperatures, then there is a temperature range where the chiral
correlation length dominates while at the lowest accessible
temperatures the spin correlation again dominates and seems to
diverge like $1/T$.
For zero temperature this field corresponds to a magnetization
$m\approx 0.15\approx 0.3 m_{\rm sat}$
(Fig.~\ref{J1J2.finite_field.fig1}(b)) where $m_{\rm sat}=0.5$ is the
saturation magnetization. Comparing with the zero temperature phase
diagrams obtained in Refs.~\onlinecite{HikiharaKecke,SudanLuescher} we
see that for these values we are in the SDW$_2$ phase but very close
to the phase boundary with the chiral phase. Our results basically
seem to confirm this picture although the oscillations $k$ apparently
slightly deviate from the value expected for a $n=2$ multipolar phase
even at the lowest temperatures (see
Fig.~\ref{J1J2.finite_field.fig1}(a)).  Our data also show that in a
certain temperature window chiral correlations can still be dominant.
This might be relevant once interchain couplings are taken into
account and might lead to a chiral phase stable at intermediate
temperatures while the spins are collinear at higher and lower
temperatures. Two magnetic phase transitions have indeed been observed
in LiCu$_2$O$_2$ where first a sinusoidal magnetic order is
established followed by a helical order at lower
temperatures.\cite{ParkChoi,RusydiMahns} While the magnetic structure
even in the low temperature phase apparently is much more complicated
than a simple planar spiral\cite{RusydiMahns} and the $J_1$-$J_2$
model does not seem to capture the essential physics of this compound
(see next section) it is nevertheless interesting that even in this
very simple model phases might be stable only in a certain temperature
range thus possibly giving rise to multiple magnetic phase
transitions once interchain couplings are taken into account.

\begin{figure}
\includegraphics*[width=0.99\columnwidth]{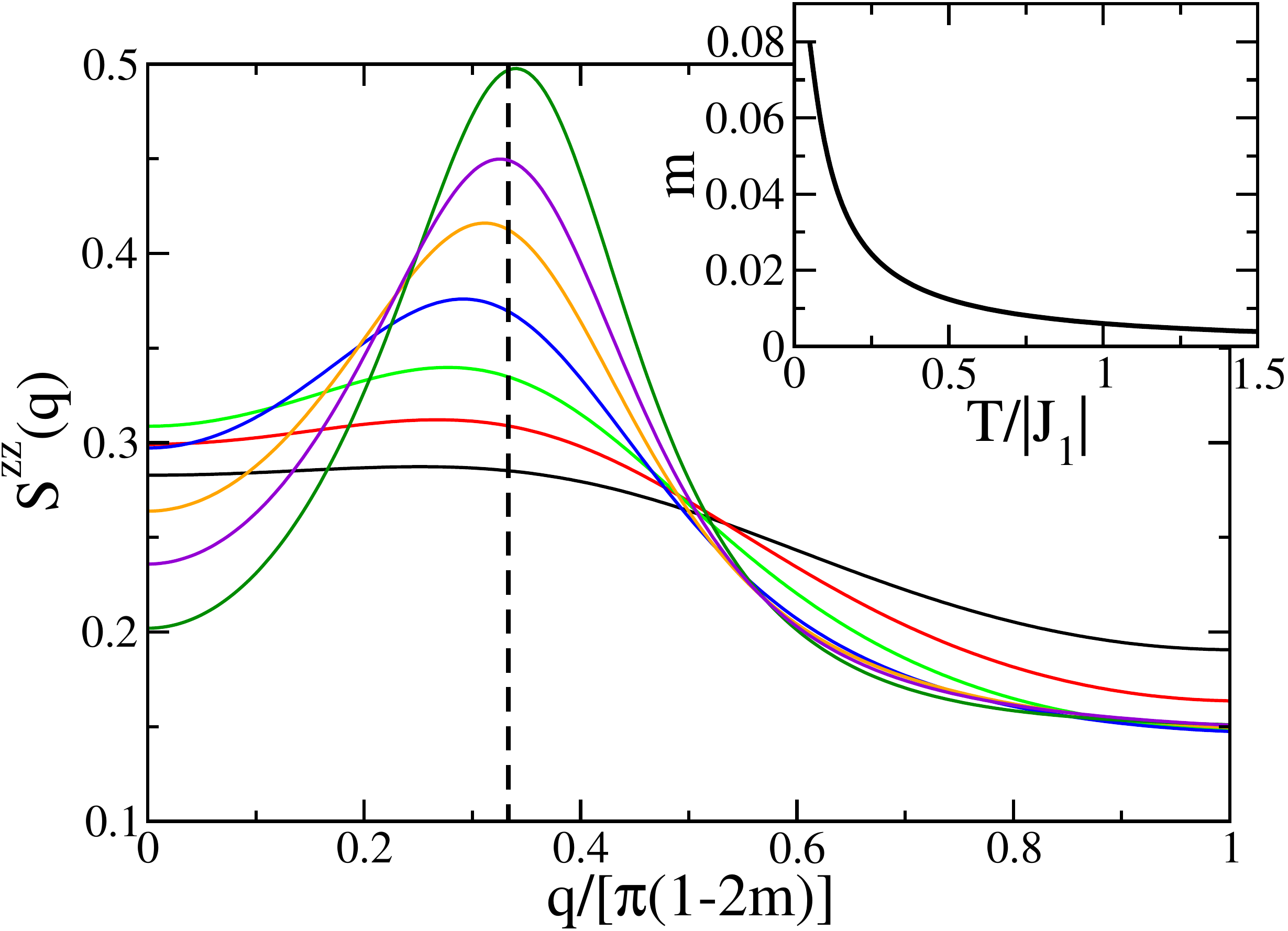}
\caption{(color online) $S^{zz}(q)$ for $\alpha=0.32$, $h=0.02 J_1$ and
  temperatures $T/|J_1| = 2.0, 1.0, 0.5, 0.2, 0.1, 0.07, 0.05$ (from botom to
  top). The dashed line marks $q/[\pi(1-2m)]=1/3$. The inset shows the
  magnetization $m$ as function of temperature. }
\label{J1J2.finite_field.fig2}
\end{figure}

In Fig.~\ref{J1J2.finite_field.fig2} the longitudinal spin structure factor
$S^{zz}(q)$ is shown for $\alpha=0.32$ and $h=0.02 J_1$.
At low temperatures $S^{zz}(q)$ is peaked at $q=\pi(1-2m)/3$ (dashed
line in Fig.~\ref{J1J2.finite_field.fig2}) indicating an $n=3$
multipolar phase. Thus experimentally a multipolar phase can already
be indentified at finite temperatures by studying the evolution of the
structure factor as obtained in neutron scattering. Such an experiment
would be most interesting for a compound close to the quantum critical
point $\alpha_c=1/4$ where small magnetic fields are sufficient to
stabilize $n=3$ or even $n=4$ multipolar phases.\cite{SudanLuescher}
As discussed in detail in Ref.~\onlinecite{DrechslerVolkova} and in
the following section, Li$_2$ZrCuO$_4$ seems to be a very promising
candidate for such a material.

\section{Multiferroic spin chain materials}
\label{Materials}
While in the corner sharing copper-oxygen chain compounds like Sr$_2$CuO$_3$
and SrCuO$_2$ the antiferromagnetic exchange interaction is of very similar
magnitude, $J\sim 2000$ K, a wide range of exchange parameters $J_1$ and $J_2$
can be found in the literature for the edge sharing compounds. For LiCuVO$_4$,
for example, fits of the susceptibility and of neutron scattering data have led
to the estimate $J_1\sim -20$ K and $J_2\sim 50$ K so that $\alpha \sim
2.5$.\cite{EnderleMukherjee} For Li$_2$ZrCuO$_4$, on the other hand,
susceptibility and specific heat data have been fitted by using $J_1\sim -300$
K and $\alpha\sim 0.3$.\cite{DrechslerVolkova} Given that the chains in these
compounds consist of the same edge sharing copper-oxygen plaquettes, this huge
variation in the magnitude of the exchange couplings is surprising.

Here we want to reanalyze the data of three of the best studied multiferroic
chain compounds namely Li$_2$ZrCuO$_4$, LiCuVO$_4$ and Li$_2$CuO$_2$. We will
concentrate on fitting susceptibility data using
\begin{equation}
\label{Mat_eq1}
\chi_{\rm exp} = \chi_0 + \chi_{J_1-J_2} \; .
\end{equation}
Here $\chi_{\rm exp}$ is the experimentally measured susceptibility,
$\chi_0$ is a constant contribution due to core diamagnetism and
Van-Vleck paramagnetism and $\chi_{J_1-J_2}$ is the numerically
calculated susceptibility for the $J_1$-$J_2$ model (\ref{intro1}).
Such fits work extremely well for the corner sharing compounds because
the intrachain coupling is about three orders of magnitude smaller
than the interchain
coupling.\cite{MotoyamaEisaki,egg94,SirkerLaflorencie,SirkerLaflorencie2,SirkerLaflorencieEPL}
For LiCuVO$_4$ it has been reported that a three-dimensional (3D)
magnetic order becomes established below $T_{3D}\sim 2.3$
K.\cite{BuettgenKrugvonNidda} While this points to intrachain
couplings which are only one or at most two orders of magnitude
smaller than the interchain couplings, a purely one-dimensional model
is still expected to be a good approximation as long as $T\gg T_{3D}$.
In Li$_2$ZrCuO$_4$ the situation has not fully been clarified yet. A
possible phase transition might occur at $T\sim 6$
K.\cite{DrechslerVolkova} Even if this is a 3D ordering transition, a
one-dimensional model should again be valid over a wide temperature
range. Therefore the expectation is that the physics of LiCuVO$_4$ and
Li$_2$ZrCuO$_4$ can largely be understood within the framework of the
$J_1$-$J_2$ model with the spin-current mechanism being responsible
for the multiferroic properties.

The situation for LiCu$_2$O$_2$, on the other hand, seems to be much more
involved.\cite{MasudaZheludev,ParkChoi,RusydiMahns,SekiYamasaki,HuangHuang}
Despite a number of studies, the magnetic structure remains
controversial. For instance, both a spiral spin order in the $ab$ as
well as in the $bc$ plane have been
reported.\cite{MasudaZheludev,SekiYamasaki} If the spin current model
is the correct explanation for the observed polarization along the
$c$-axis\cite{ParkChoi} then the $bc$ spiral must be realized. The two
magnetic ordering transitions at comparitatively large temperatures,
$T\approx 22$ K and $T\approx 24$ K,\cite{RusydiMahns} point to much
larger interchain couplings than for the two other compounds discussed
above. In fact, it has been found that there are substantial spin
correlations not only along the chain direction ($b$-axis) but also in
the plane of the copper-oxygen plaquettes perpendicular to the chain
direction ($a$-axis) making the compound at low temperatures almost
two dimensional.\cite{HuangHuang} One might speculate that the reason
why the magnetic properties of this material are so different is
related to the fact that two different copper sites exist - the
in-chain Cu$^{2+}$ and the Cu$^{1+}$ interconnecting different chains.
This might lead to substantial charge fluctuations and thus to
enhanced magnetic interchain couplings. Nevertheless, a fit of the
susceptibility at high temperatures using the $J_1$-$J_2$ model might
still be useful to obtain an estimate for the magnitude of the $J_1$
and $J_2$ couplings.

The absolute values of the measured maxima of $\chi$ already allow some
general statements about the magnitude of the exchange constants and the
frustration parameters $\alpha$. In LiCuVO$_4$ and in LiCu$_2$O$_2$ the
susceptibility is about two orders of magnitude larger than in the corner
sharing chain compounds Sr$_2$CuO$_3$ and SrCuO$_2$ so that the
antiferromagnetic exchange constant $J_2$ should also roughly be two orders of
magnitude smaller. In Li$_2$ZrCuO$_4$ the maximum is almost an order larger
than in the other two compounds clearly indicating that this compound should
be closer to the quantum critical point $\alpha_c=1/4$ than the other two
compounds. If the magnetic exchange constants are of comparable magnitude,
then $\alpha$ is expected to be largest for LiCu$_2$O$_2$ and smallest for
Li$_2$ZrCuO$_4$ with LiCuVO$_4$ having an intermediate frustration parameter.

\begin{figure}
\includegraphics*[width=0.99\columnwidth]{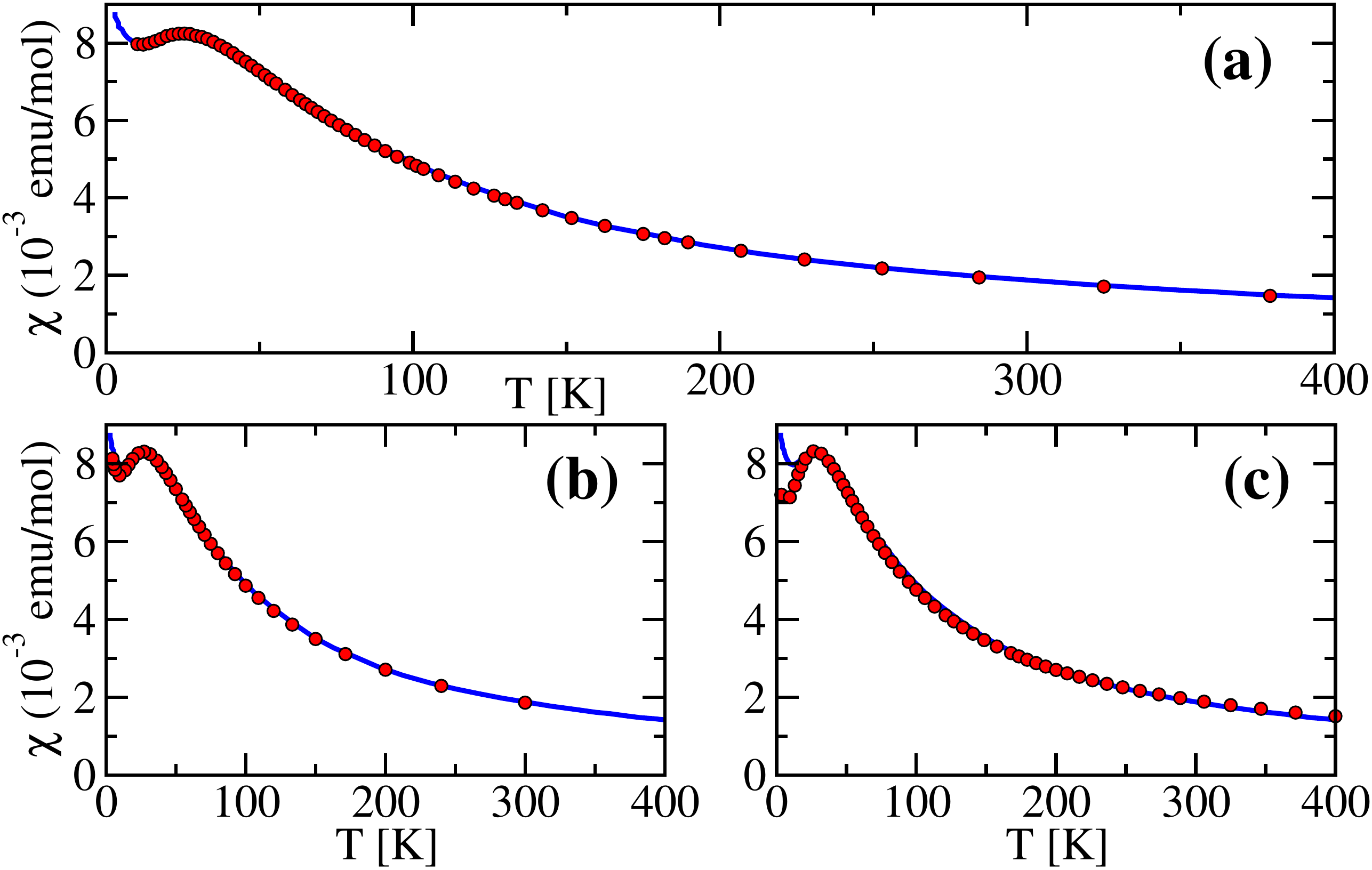}
\caption{(color online) The experimentally measured susceptibility of
  LiCuVO$_4$ for $H\parallel c$ (solid line) compared to fits (dots)
  following Eq.~(\ref{Mat_eq1}) with a gyromagnetic ratio
  $g=2.313$.\cite{BuettgenKrugvonNidda,KrugvonNiddaSvistov} (a) The
  best fit is obtained with $\chi_0=6\cdot 10^{-5}$ emu/mol, $J_2=91$
  K, and $\alpha=0.5$. (b) An alternative fit with $\chi_0=1.4\cdot
  10^{-4}$ emu/mol, $J_2=72$ K, and $\alpha=0.6$. (c) An exchange
  coupling $J_2=52$ K similar to the one in
  Ref.~\onlinecite{EnderleMukherjee} can be used with $\chi_0=2.7\cdot
  10^{-4}$ emu/mol, however, the frustration parameter $\alpha=1.0$
  chosen to obtain the best fit is still much smaller than the one in
  Ref.~\onlinecite{EnderleMukherjee} and the fit is much worse than
  the ones shown in (a) and (b).}
\label{J1J2.fig_mat1}
\end{figure}

The susceptibility of LiCuVO$_4$ shown in Fig.~\ref{J1J2.fig_mat1} has
a maximum and a local minimum at low temperatures. If $\chi$ can
indeed be described by the $J_1$-$J_2$ model then a comparison with
Fig.~\ref{J1J2.fig1} indicates that $\alpha\sim 0.5$. The large
$\alpha$ used in Ref.~\onlinecite{EnderleMukherjee} can certainly not
explain this structure. By performing a fit according to
(\ref{Mat_eq1}) we find that $J_2=91$ K and $\alpha=0.5$ yields the
best fit as shown in Fig.~\ref{J1J2.fig1}(a). A good fit is also
possible with $J_2=72$ K and $\alpha=0.6$ (Fig.~\ref{J1J2.fig1}(b)).
If we reduce the next-nearest neighbor exchange to $J_2\sim 50$ K as
assumed in Ref.~\onlinecite{EnderleMukherjee} we have to choose
$\alpha\sim 1$ to obtain the best fit possible with this value of
$J_2$ (Fig.~\ref{J1J2.fig1}(c)). However, such a fit fails to
reproduce the low temperature structure.

\begin{figure}
\includegraphics*[width=0.99\columnwidth]{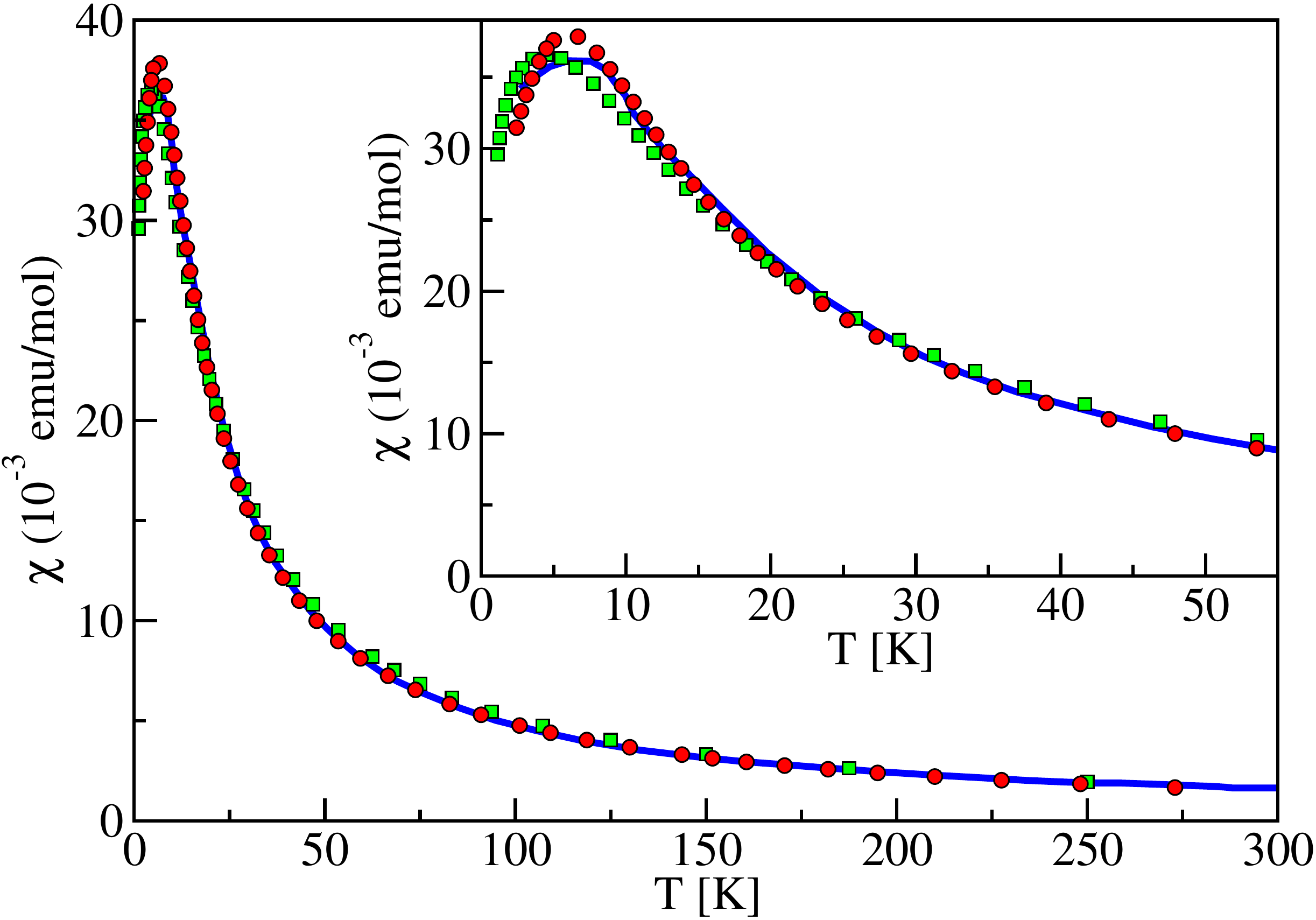}
\caption{(color online) The measured susceptibility (solid line) for
  Li$_2$ZrCuO$_4$ taken from Ref.~\onlinecite{DrechslerVolkova}
  compared to fits using the $J_1-J_2$ model. An excellent fit
  (circles) is obtained using $\alpha=0.3$, $g=2$ and $J_1=-273$ K
  ($J_2=81.9$ K) with $\chi_0=0$ confirming the analysis in
  Ref.~\onlinecite{DrechslerVolkova} based on exact diagonalization
  data. A reasonable fit (squares) is also possible with $\alpha=0.4$,
  $g=2.2$, $\chi_0=0$, and $J_1=-75$ K. Inset: Blow-up of the low
  temperature region.}
\label{J1J2.fig_mat2}
\end{figure}

The susceptibility data for Li$_2$ZrCuO$_4$ have already been analyzed
in Ref.~\onlinecite{DrechslerVolkova} by comparing with exact
diagonalization data for small rings consisting of up to $N=20$ sites.
Finite size effects are expected to be neglegible if $T\gg v/N$ where
$v$ is the spin velocity. The spin velocity is of the order of the
exchange constant so that the finite size data for $\chi$ should be
reliable for $T\gtrsim 20$ K. Given that the maximum of $\chi$ is
located at $T\approx 7.6$ K, i.e., at temperatures where finite size
effects might play a role, it is helpful to reanalyze the experimental
data using the TMRG algorithm which yields results directly for the
infinite system. As shown in Fig.~\ref{J1J2.fig_mat2} very similar
fits to the one in Ref.~\onlinecite{DrechslerVolkova} are obtained.
While the magnitude of the frustration parameter can only be varied slightly
if one wants to obtain a reasonable fit, the exchange parameters change
dramatically, for example, from $J_1=-273$ K for $\alpha=0.3$ to $J_1=-75$ K
for $\alpha=0.4$. In LDA calculations $J_1=-151\pm 35$ K and $J_2=35\pm 12$ K
was found which also will allow for a reasonable fit with a frustration
parameter $\alpha\in [0.3,0.4]$.\cite{DrechslerVolkova} We therefore confirm
the main conclusion of Ref.~\onlinecite{DrechslerVolkova} that Li$_2$ZrCuO$_4$
is close to the critical point $\alpha_c=1/4$.

\begin{figure}
\includegraphics*[width=0.99\columnwidth]{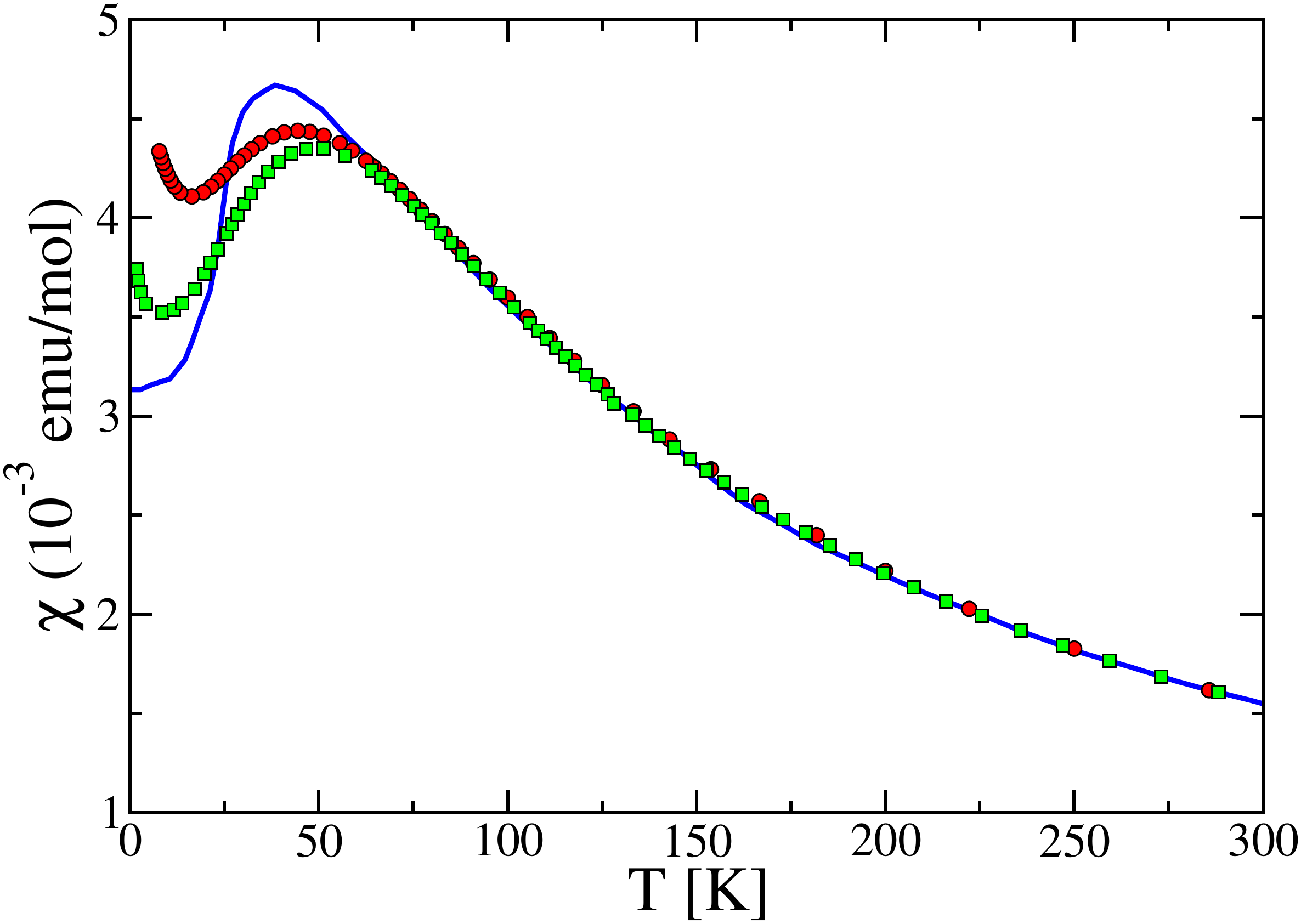}
\caption{(color online) The measured susceptibility (solid line) for
  LiCu$_2$O$_2$ with $H\parallel c$ taken from
  Ref.~\onlinecite{MasudaZheludev} compared to fits using the
  $J_1-J_2$ model. With $g=2.2$ the best fit is obtained with
  $\alpha=0.6$, $J_2=120$ K and $\chi_0=0$ (circles) while for $g=2.3$
  a fit with $\alpha=1.6$, $J_2=83$ K and $\chi_0=0$ works best
  (squares).}
\label{J1J2.fig_mat3}
\end{figure}

Finally, we want to analyze the susceptibility data for LiCu$_2$O$_2$. As
already mentioned, we do not expect that the $J_1$-$J_2$ model will describe
the susceptibility of this compound at low temperatures and indeed we find
that it is impossible to obtain a good fit down to low temperatures (see
Fig.~\ref{J1J2.fig_mat3}). 
If we concentrate on temperatures large compared to the magnetic
ordering transitions at $T\sim 20$ K, then a fit is possible and we
find that $J_2\approx 80-120$ K. Because the susceptibility can only
be fitted at high temperatures, the frustration parameter cannot be
fixed precisely.

It is, however, important to note that the best fits for all three
compounds yield values of $J_2\lesssim 100$ K while $J_1\sim -200$ K.
This confirms our expectation that these materials consisting of the
same edge sharing copper-oxygen plaquettes should have very similar
magnetic exchange constants as is also the case for the corner sharing
chain compounds. The values we obtain from the susceptibility fits are
consistent with values in the literature for LiCu$_2$O$_2$ ($J_1\sim
-120$ K, $J_2\sim 80$ K) \cite{MasudaZheludev,ParkChoi} and for
Li$_2$ZrCuO$_4$.\cite{DrechslerVolkova} For LiCuVO$_4$, however, we
find values which differ dramatically from the values given in
Ref.~\onlinecite{EnderleMukherjee} which later were also used in a
number of other publications. We want to stress again that these
values are not consistent with the latest susceptibility
data.\cite{BuettgenKrugvonNidda} An analysis of neutron scattering
data using a standard spin-wave dispersion which seems to confirm
these values \cite{EnderleMukherjee} is in our opinion not applicable
here.  In such an analysis the magnon bandwidth is directly determined
by the bare exchange couplings while the stark deviation of the
quantum from the classical pitch angle (see Fig.~\ref{J1J2.fig4}(b))
suggests that the frustration parameter is strongly renormalized due
to quantum fluctuations.

The pitch angle measured experimentally in
Ref.~\onlinecite{EnderleMukherjee} of $\phi\approx 83.7^\circ$ is in
fact in excellent agreement with the frustration parameters
$\alpha=0.5-0.6$ obtained from the fits in Fig.~\ref{J1J2.fig_mat1}.
According to Fig.~\ref{J1J2.fig4}(b) we have a quantum pitch angle
$\phi\approx 83.5^\circ$ for $\alpha=0.5$ and $\phi\approx 87.3^\circ$
for $\alpha=0.6$. It is important to stress again that only in the
classical model large frustration parameters are needed to obtain
pitch angles close to $90^\circ$. We expect, according to
Fig.~\ref{J1J2.fig4}(a), that the pitch angle in LiCuVO$_4$ can be
reduced by $15-20\%$ by increasing the temperature to $T\sim J_2\sim
90$ K. For Li$_2$ZrCuO$_4$ the magnetic structure has not been studied
so far. The pitch angle for $\alpha=0.3$ at zero temperature obtained
from our numerical calculations is $\phi\approx 51.5^\circ$. Here we
expect a large variation with temperature (see Fig.~\ref{J1J2.fig3}
and Fig.~\ref{J1J2.fig6}(b)) and it would be interesting to see if
this can also be observed experimentally. For LiCu$_2$O$_2$ a pitch
angle $\phi\approx 62.6^\circ$ has been
measured.\cite{MasudaZheludev,ParkChoi} Such a small pitch angle
cannot be explained within the $J_1$-$J_2$ model given that
$\alpha\gtrsim 0.6$ according to the fits shown in
Fig.~\ref{J1J2.fig_mat3}. We are therefore again lead to the
conclusion that the $J_1$-$J_2$ model cannot explain the experimental
data for this compound.

\section{Conclusions}
\label{conc}
The rich physics of the $J_1$-$J_2$ model with ferromagnetic nearest
neighbor coupling $J_1$ and antiferromagnetic next-nearest neighbor
coupling $J_2$ has attracted a lot of interest recently. The phase
diagram for this model including exchange anisotropies and magnetic
fields has been addressed in a number of numerical
studies.\cite{BursillGehring,FurukawaSato,HeidrichMeisnerMcCulloch,HikiharaKecke,SudanLuescher}
Here we have shown that the physical properties of this simple model
are even more intriguing if the interplay of quantum and thermal
fluctuations is taken into account.  In particular, we found that the
incommensurate oscillations of the spin-spin correlation function (the
quantum analogue of the pitch angle of the classical spiral order)
does not only strongly depend on the frustration parameter
$\alpha=J_2/|J_1|$ but also on temperature. For zero temperature we
find an incommensurability (pitch angle) $\phi$ for $\alpha\gtrsim
0.6$ which is very close to $90^\circ$ in the quantum model and thus
very different from the classical pitch angle
$\phi=\arccos(1/4\alpha)$ in agreement with an earlier
study.\cite{BursillGehring} At temperatures $T\sim J_2$, however, the
pitch angle is much closer to the classical value for $\alpha\gtrsim
0.5$. Furthermore, we find a very strong temperature dependence of the
pitch angle for frustration parameters close to the critical point
$\alpha_c=1/4$. We therefore expect that the wave vector where the
static spin structure factor for Li$_2$ZrCuO$_4$ is peaked - a
compound which according to Ref.~\onlinecite{DrechslerVolkova} and the
susceptibility analysis performed here has a frustration parameter
$\alpha\sim 0.3-0.4$ - varies significantly with temperature. For this
range of frustration parameters we find that a small easy-plane
anisotropy (at zero magnetic field) leads to long-range chiral order
even in the purely one-dimensional model.\footnote{Long-range chiral
  order has also recently been found numerically in
  Refs.\onlinecite{HikiharaKecke,HeidrichMeisnerMcCulloch} for finite
  magnetizations.} We could, however, find no evidence for the dimer
phase which was predicted in Ref.~\onlinecite{SommaAligia} for the
anisotropic case for larger frustration parameters. Another
observation which again might be relevant for future studies on
Li$_2$ZrCuO$_4$ is that small magnetic fields (which would correspond
to $5-7$ T for Li$_2$ZrCuO$_4$) can stabilize an SDW$_3$ ($n=3$
multipolar) phase for $\alpha\sim 0.3$. We suggest that such a phase
can already be identified by neutron scattering at finite temperatures
by monitoring at the same time the peak position $q_{\rm max}$ of the
static longitudinal structure factor and the magnetization $m$. The
signature of the SDW$_3$ phase would be that $q_{\rm max}\to
\pi(1-2m)/3$ for $T\to 0$.

We also reanalyzed the susceptibility data for LiCuVO$_4$. Here the
newer data in Ref.~\onlinecite{BuettgenKrugvonNidda} seem to be of
much better quality than the older data in
Ref.~\onlinecite{EnderleMukherjee}. The newer data clearly show that
the exchange parameters $J_1\sim -20$ K, $J_2\sim 50$ K and
$\alpha=2.5$ found in Ref.~\onlinecite{EnderleMukherjee} do not yield
a reasonable fit. These exchange parameters also seem very unlikely
given that they deviate significantly from those in other edge-sharing
copper-oxygen chains. Our analysis shows that the susceptibility data
are best fitted with $J_2\approx 70-90$ K and $\alpha=0.5-0.6$.

In LiCu$_2$O$_2$ two magnetic ordering transitions already occur at
temperatures $T_{3D}\sim 20$ K and an analysis based on the purely
one-dimensional $J_1$-$J_2$ model clearly cannot be as successful as
for LiCuVO$_4$ and Li$_2$ZrCuO$_4$. For $T\gg T_{3D}$ we have shown
that a reasonable fit of the susceptibility is nevertheless possible
leading to $J_2\sim 80-120$ K. The frustration parameter remains,
however, ambiguous in such a high temperature fit and we find
$\alpha\in [0.6,1.6]$ with the best fit being obtained for
$\alpha=0.6$.

Based on the best fits of the susceptibilities we conclude that all
three considered compounds seem to have very similar exchange
constants $J_1\sim -200$ K and $J_2\lesssim 100$ K. The frustration
parameter, on the other hand, varies from $\alpha\approx 0.3$ for
Li$_2$ZrCuO$_4$, $\alpha=0.5-0.6$ for LiCuVO$_4$ to $\alpha\gtrsim
0.6$ for LiCu$_2$O$_2$. The parameters found here for LiCuVO$_4$ are
fully consistent with the measured pitch angle $\sim 84^\circ$. For
LiCu$_2$O$_2$, on the other hand, the small measured pitch angle $\sim
63^\circ$ cannot be explained within the $J_1$-$J_2$ model stressing
again that this model is not sufficient to explain the experimental
data for this compound.  The smallest pitch angle is expected for
Li$_2$ZrCuO$_4$. Based on our numerical calculations we predict
$\phi\sim 53^\circ$.

\begin{acknowledgments}
  JS thanks I. McCulloch, S. Drechsler, A. Furusaki, P. Horsch, and J.
  Richter for valuable discussions.
\end{acknowledgments}


\end{document}